\documentclass[aps,prx,twocolumn,floatfix,superscriptaddress,nofootinbib]{revtex4-2}

\usepackage{amsmath,amsthm,amssymb}
\usepackage{mathtools}
\usepackage{dsfont}
\usepackage{physics,color,comment}
\usepackage{wasysym}
\usepackage{subcaption}
\usepackage{graphicx}
\usepackage{dcolumn}
\usepackage{bm}
\usepackage[hidelinks]{hyperref}
\usepackage[capitalize]{cleveref}
\usepackage{gensymb}

\newcommand{\expv}[1]{\langle #1 \rangle}

\newcommand{\bea}{\begin{eqnarray}}
\newcommand{\eea}{\end{eqnarray}}
\newcommand{\eq}[1]{Eq.~(\ref{#1})} %
\newcommand{\fig}[1]{Fig.~\ref{#1}} %

\newcommand{\mcs}{\mathcal{S}}
\newcommand{\sfull}{\mathcal{S}_{\mathrm{full}}}

\newcommand{\mci}{\mathcal{I}}
\newcommand{\ifull}{\mathcal{I}_{\mathrm{full}}}
\newcommand{\mco}{\mathcal{O}}
\newcommand{\mcp}{P}

\begin{document}
\title{Bounding Eigenstate Overlap from Hamiltonian Moments: Success Probability Guarantees for Quantum Phase Estimation}
\author{Junan Lin}
\affiliation{National Research Council Canada, Toronto, ON, Canada}
\author{Artur F. Izmaylov}
\email{artur.izmaylov@utoronto.ca}
\affiliation{Chemical Physics Theory Group, Department of Chemistry, University of Toronto, Toronto, Ontario M5S 3H6, Canada}
\affiliation{Department of Physical and Environmental Sciences, University of Toronto Scarborough, Toronto, Ontario M1C 1A4, Canada}

\begin{abstract}
Estimating the overlap between a prepared state and a target eigenstate is crucial for the efficiency of quantum phase estimation (QPE), since QPE succeeds with probability equal to this overlap. We present a systematically improvable method to compute certified upper and lower bounds on such overlaps using a finite set of Hamiltonian moments. Our approach constructs optimal polynomial upper/lower bounds on an energy-window indicator and evaluates them through linear and semidefinite programs, yielding the tightest bounds consistent with the available moment and spectral-interval information. We demonstrate the method on strongly correlated molecular Hamiltonians and study the impact of approximate moments obtained from tensor-network contractions. The resulting bounds provide a practical pre-QPE screening tool for selecting initial states and can be implemented with either classical moment computation or quantum expectation estimation.
\end{abstract}
\pacs{}
\maketitle

\section{Introduction}
Quantum phase estimation (QPE) is a core primitive for many quantum algorithms~\cite{Kitaev1995,Abrams1999,Aspuru-Guzik2005,Ge2019,Moore2021a,Lin2022,Wang2023},
encoding Hamiltonian eigenvalues as phases of a unitary operator.
In QPE, the probability of sampling an eigenvalue is set by the overlap of the input state with the corresponding Hamiltonian eigenstate(s); when this overlap is small, QPE-based workflows can become impractical.
Thus, QPE performance hinges on (i) implementing Hamiltonian evolution (or equivalent encodings) with feasible resources and (ii) preparing an initial state with sufficient overlap with the target eigenstate(s).

Recent work has improved Hamiltonian encoding via Trotterization~\cite{Lloyd1996,Kivlichan2020,Childs2021,Martinez-Martinez2023,martinez2023En},
Linear Combination of Unitaries (LCU)~\cite{Childs2012,An2023,Loaiza2024,loaiza2023BLISS,loaiza2023LCU,Patel2024,caesura2025LCU,deka2024SS},
and qubitization~\cite{Low2019,Berry2019,Lee2021,low2025sqrt},
as well as state preparation methods such as coarse-QPE refinement~\cite{Fomichev2024},
matrix product state approximation~\cite{Berry2024}, orbital optimization~\cite{Ollitrault2024}, and quantum embedding~\cite{Erakovic2024},
motivated by concerns that state preparation may become a bottleneck in some QPE-based workflows~\cite{Lee2023}.

A key question is when QPE offers a quantum advantage over classical methods.
When the initial state is constructed classically, QPE is most useful in an intermediate regime: classical methods can produce a state with non-negligible overlap with the target eigenstate, but not the desired energy accuracy.
Empirically, for small strongly correlated molecules, achieving $>50\%$ overlap can be easier than reaching chemical accuracy~\cite{Choi2024}, and DMRG studies show that significant overlap often requires lower tensor ranks than chemical accuracy~\cite{Berry2024}.
This motivates a practical screening task: before committing to deep QPE circuits, certify whether a candidate state has sufficiently large overlap (success probability) to justify the QPE cost.

Ideally, one would assess the overlap before running QPE, but general-purpose methods for \emph{certifying} bounds on $P_i$ remain limited.
Coherent approaches can estimate overlaps using spectral filtering combined with amplitude estimation, or by post-processing QPE outcomes; however, they require a coherent ``good-state'' marking oracle (typically implemented via QPE- or QSVT-based filtering/projectors) with precision set by the required spectral resolution and gaps, which can demand substantial circuit depth and high precision.
In contrast, many workflows naturally provide access to low-order Hamiltonian moments, either classically (e.g., from tensor-network contractions) or via expectation-estimation primitives on quantum devices.
This leads to a complementary question for pre-QPE screening:
\emph{given only a finite set of Hamiltonian moments and coarse spectral information, what overlap bounds can be certified?}

Precursors to moment-based overlap bounds date back to Eckart~\cite{Eckart1930}, who derived the ground-state lower bound
\begin{equation}\label{eqn:Eckart_bound}
    \left |\bra{\phi} E_0\rangle \right|^2 \geq \frac{E_{1}-\expv{\hat H}}{E_{1} - E_{0}},
\end{equation}
where $E_0$ and $E_1$ are the ground and first excited eigenvalues of $\hat H$, and $\expv{\hat H} = \bra{\phi}\hat H\ket{\phi}$.
This bound requires exact energies, applies only to the ground state, and becomes uninformative when $\expv{\hat H}>E_1$ (a common issue in strongly correlated settings~\cite{Choi2024}).
More recently, Mora \emph{et al.}~\cite{Mora:PRL/2007} proposed the upper bound
\begin{equation}\label{eqn:Louvet_bound}
    \left |\bra{\phi} E_0\rangle \right|^2 \leq \frac{(\langle \hat H \rangle - E_0)^2}{2 (\langle \hat H^2 \rangle - \langle \hat H \rangle^2)}.
\end{equation}
Upper bounds are particularly useful for screening because if the certified upper bound falls below a threshold, the state can be excluded~\cite{louvet2023gono}.
This bound is again ground-state-specific, depends on the exact ground-state energy, and it is unclear whether it is optimal among all bounds based only on the first two Hamiltonian moments.

In this work, we develop a unified framework for \emph{certifying} upper \emph{and} lower bounds on overlaps of an approximate state with selected eigenstates, degenerate eigenspaces, or unresolved spectral regions.
Our key contribution is to cast overlap certification as a truncated moment problem and compute the \emph{tightest possible} bounds consistent with (i) a finite set of Hamiltonian moments $\expv{\hat H^n}$ and (ii) coarse spectral information in the form of eigenvalue interval enclosures (with exact eigenvalues as a limiting case).
The resulting bounds are obtained via linear and semidefinite programs and are optimal in the sense that no strictly tighter bounds can be inferred from the same information.

The rest of the paper is organized as follows.
Section~II develops the optimization formulations (LP/SIP/SDP), establishes optimality via primal--dual convex duality, and presents numerical demonstrations on model systems and strongly correlated molecular Hamiltonians, including approximate moments from tensor-network contractions.
Section~III concludes and provides the outlook.

\section{Theory}
\subsection{Problem setup and moment information}

Let $\hat H = \sum_{i=0}^{D} E_i \ketbra{E_i}{E_i}$ be the spectral decomposition of the Hamiltonian.
We first present the algorithm when the $E_i$'s are known exactly as a useful limiting case, then a more useful version when only interval enclosures enclosing certain desirable eigenvalues are known. 
We also assume throughout this work that $\hat{H}$ has a normalized spectrum in the range $[-1,1]$ for simplicity: if this does not hold, we can apply the linear rescaling
\begin{equation}\label{eqn:rescaled_H}
    \hat{H}_{\text{rs}} = \frac{2 \hat{H} - (E_L + E_{U})I}{E_{U} - E_L},
\end{equation}
where $I$ is the identity operator, and $E_L$ ($E_U$) is a lower (upper) spectral bound used for rescaling.
As long as $E_L \leq E_0$ and $E_U \geq E_D$, the spectrum of $\hat H_{\text{rs}}$ lies inside $[-1,1]$.

Denote the vector of overlaps between an input (trial) state $\ket{\phi}$ and each eigenstate $\ket{E_i}$ as $P$, with components
\begin{equation}
    P_i = \left |\bra{\phi} E_i\rangle \right|^2,\qquad \sum_{i=0}^{D} P_i=1,\qquad P_i \geq 0.
\end{equation}
Our goal is to obtain certified upper and lower bounds on a sum of selected overlaps,
\begin{equation}\label{eqn_target_val_def}
    \mcp_{\mcs} = \sum_{i \in \mcs} P_i,
\end{equation}
where $\mcs \subset \sfull = \{0,1,\dots, D\}$ is the index set specifying the eigenstates of interest (e.g., $\mcs=\{0\}$ for the ground state).

It is convenient to write $\mcp_{\mcs}$ using the indicator function $f_{\mcs}(E)$,
\begin{equation}
    \mcp_\mcs = \sum_{i \in \sfull} P_i f_\mcs(E_i)=f^\top P,\quad
    f_\mcs(E_i) = \begin{cases}
        1,\ i \in \mcs \\
        0,\ i \notin \mcs,
    \end{cases}
\end{equation}
where we slightly abuse notation and use $f$ to denote the vector with entries $f_i=f_{\mcs}(E_i)$.
In principle, $f_\mcs(E)$ can be defined arbitrarily away from the eigenvalues.
Using the functional calculus, we can also write
\begin{equation}
    \mcp_\mcs = \bra{\phi} f_\mcs(\hat H) \ket{\phi} \coloneqq \expv{f_\mcs(\hat H)}_{\phi}.
\end{equation}
Directly evaluating $\expv{f_\mcs(\hat H)}_{\phi}$ is infeasible because it requires knowledge of both $\{E_i\}$ and $\{P_i\}$.

\paragraph*{Chebyshev moments.}
Let $T_n$ be the Chebyshev polynomials of the first kind.
We assume access to the \emph{Chebyshev moments}
\begin{equation}\label{eqn:cheb_moments_def}
    M_n \coloneqq \expv{T_n(\hat H_{\rm rs})}_{\phi},\qquad n=0,1,\dots,N,
\end{equation}
and collect them into the vector $M=(M_0,\dots,M_N)^\top$.
If one instead computes power moments $\expv{\hat H^n}$, the Chebyshev moments can be obtained by a fixed linear transformation; we use the Chebyshev form throughout for numerical stability on $[-1,1]$.

For any degree-$N$ polynomial written in the Chebyshev basis,
\begin{equation}\label{eqn:p_H}
    p_{N}(\hat H_{\rm rs}) = \sum_{n=0}^{N} c_n T_n(\hat{H}_{\rm rs}),
\end{equation}
its expectation value depends \emph{linearly} on the moments:
\begin{equation}\label{eqn:poly_expect}
    \expv{p_{N}(\hat H_{\rm rs})}_{\phi} = \sum_{n=0}^{N} c_n \expv{T_n(\hat{H}_{\rm rs})}_{\phi}
    = c^\top M.
\end{equation}
Thus, once $M$ is available, evaluating $\expv{p_N(\hat H_{\rm rs})}_{\phi}$ reduces to an inner product.

\subsection{Special case: Using exact eigenvalues}\label{subsec_exact_EV}
Before turning to the more realistic setting of interval eigenvalue information, we first consider the limiting case where the eigenvalues $\{E_i\}$ are known exactly.
The remaining design problem is: \emph{how should the coefficients $c_n$ be chosen so that $p_N$ provably bounds the indicator while being as tight as possible?}
This exact-eigenvalue case can be viewed as a systematically improvable extension of Eckart-type variational overlap bounds: it uses higher-order moment information to produce optimal certified lower and upper bounds, and it applies to arbitrary target sets (not only the ground state).

We focus on the upper bound (denoted by superscript $+$); the lower bound (denoted by superscript $-$) is obtained by swapping majorization with minorization, $\min$ with $\max$, and $\ge$ with $\le$.
To obtain an upper bound, we consider polynomials $p_N(E)$ that \emph{majorize} $f_{\mcs}$ at the eigenvalues:
\begin{equation}\label{eqn_majorize_condition}
p_{N}^{+}(E_i) \ge f_{\mcs}(E_i)\quad \forall i \in \sfull.
\end{equation}
Any such polynomial yields an upper bound because
\begin{equation}
\expv{p_{N}^{+}(\hat H_{\rm rs})}_{\phi}
= \sum_i P_i\, p_N^+(E_i)
\ge \sum_i P_i\, f_{\mcs}(E_i)
= \mcp_\mcs.
\end{equation}

The tightest degree-$N$ upper bound is obtained by minimizing $c^\top M$ subject to the majorization constraints.
Define the \emph{Chebyshev--Vandermonde} matrix $\mathbf{T}\in \mathbb{R}^{(D+1)\times (N+1)}$ with entries
\begin{equation}
    \mathbf{T}_{i,n} \coloneqq T_n(E_i),\qquad i=0,\dots,D,\ \ n=0,\dots,N.
\end{equation}
Then the optimal coefficients solve the linear program (LP)
\begin{equation}
\begin{aligned}
& \underset{c \in \mathbb{R}^{N+1}}{\mathrm{minimize}} & & c^\top M \\
& \text{subject to} & & \mathbf{T} c \ge f.
\end{aligned}\label{eqn_LP}
\end{equation}
We denote the solution to \eq{eqn_LP} by $\mcp_\mcs^+$.
This LP has $N+1$ variables and $D+1$ constraints; in practice it is straightforward to solve with standard solvers~\cite{Cohen2019}.
The same formulation applies whether $\mcs$ contains one index or multiple indices (total overlap with multiple eigenstates).
The exact-eigenvalue assumption will next be relaxed in \Cref{sec_approx_eval}.

\subsection{Interval overlap bounds via semidefinite programming}
\label{sec_approx_eval}
We now generalize the discrete-eigenvalue formulation to the realistic setting where exact eigenvalues are unavailable.
The key change is to replace finitely many pointwise constraints at $\{E_i\}$ with continuum constraints over energy intervals known to contain eigenvalues.

As a motivating example, suppose $E_0$ is not known exactly but is guaranteed to lie in an interval $[-1, E_0^+]$, while all higher eigenvalues lie in a non-overlapping region $[E_1^-, 1]$ with $E_1^- \geq E_0^+$.
Such enclosures can be obtained from the variational principle and energy bound theories including Temple~\cite{Temple1928}, Lehmann~\cite{Lekmann1949,Beattie1998}, and Pollak--Martinazzo~\cite{Pollak2020,Ronto2023}.
Defining an indicator that is $1$ on $[-1,E_0^+]$ and $0$ on $[E_1^-,1]$, any majorant/minorant polynomial over these regions yields certified upper/lower bounds on $P_0$ without knowing the exact eigenvalues.

We now define the interval formulation formally.
Let
\begin{equation}
    \ifull \coloneqq \bigcup_{k=1}^{K} \mci_k \subseteq [-1,1]
\end{equation}
be a union of disjoint intervals $\mci_k$ that cover the spectral range of $\hat H_{\rm rs}$ while excluding regions known to contain no eigenvalue.
Let $\mci \coloneqq \bigcup_{j \in \mathcal{J}} \mci_j \subset \ifull$ be the union of selected intervals (the target region).
Define the interval indicator
\begin{equation}
    f_\mci(E) = \begin{cases}
        1,\ E \in \mci \\
        0,\ E \notin \mci.
    \end{cases}
\end{equation}
Let $d\mu(E)=\sum_i P_i \delta(E-E_i) dE$ be the discrete spectral measure induced by $\ket{\phi}$, where $\delta(\cdot)$ is the Dirac delta function.
Then the target overlap is
\begin{equation}
    \mcp_{\mci}= \int_{\ifull} f_{\mci}(E)\, d\mu(E),
\end{equation}
i.e., the total overlap with all eigenstates whose eigenvalues lie in $\mci$.

To obtain bounds from the moments $M$, we seek polynomials that majorize/minorize $f_{\mci}$ \emph{over all energies in $\ifull$}.
Focusing on the upper bound, we require
\begin{equation}
p_{N}^{+}(E) \ge f_{\mci}(E)\quad \forall E \in \ifull,
\end{equation}
and minimize $c^\top M$ over such polynomials.
This yields the semi-infinite program (SIP)~\cite{Hettich1986,Lopez2007}
\begin{equation}\label{eq_ub_sip}
\begin{aligned}
    \underset{c \in \mathbb{R}^{N+1}}{\mathrm{minimize}} \quad &c^\top M  \\
    \text{subject to } \quad &p(E) - f_{\mci}(E) \ge 0,\ \forall E \in \ifull.
\end{aligned}
\end{equation}
We denote the solution to \eq{eq_ub_sip} by $\mcp_\mci^+$.
The lower bound is obtained by replacing $\min$ with $\max$, $\ge$ with $\le$, and $+$ with $-$.


The main difficulty in \eqref{eq_ub_sip} is enforcing nonnegativity over a continuum of energies.
For each interval $\mci_k=[l_k,r_k]$, we use the Markov--Luk\'acs certificate: a polynomial $q(E)$ of degree $N$
satisfies $q(E)\ge 0$ for all $E\in[l_k,r_k]$ if and only if it admits a \emph{sum-of-squares} (SOS) representation
\begin{equation}\label{eq:ML_cert}
q(E)=
\begin{cases}
x(E)^2 + (E-l_k)(r_k-E)\,y(E)^2, & N=2m,\\[2pt]
(E-l_k)\,z(E)^2 + (r_k-E)\, w(E)^2, & N=2m+1,
\end{cases}
\end{equation}
for some polynomials $x,y,z,w$ of appropriate degrees.
We parametrize the squares via Gram matrices, e.g.,
$x(E)^2 = v_m(E)^\top Q_x^{(k)} v_m(E)$ with $v_m(E)=(1,E,\ldots,E^m)^\top$ and $Q_x^{(k)}\succeq 0$
(and similarly for $y,z,w$). This converts nonnegativity on $\mci_k$ into semidefinite constraints.

To write the resulting constraints in standard SDP form, we enforce the identity
\begin{equation}
p(E)-f_\mci^{(k)}(E) = q_{\mathrm{nn}}^{(k)}(E)\qquad \forall E\in \mci_k,\quad k=1,\dots,K,
\end{equation}
where $f_\mci^{(k)}(E)\in\{0,1\}$ is the constant value of the indicator on interval $\mci_k$ and
$q_{\mathrm{nn}}^{(k)}$ has the Markov--Luk\'acs form \eqref{eq:ML_cert}.
Although $p_N$ is represented in the Chebyshev basis, the SOS constraints are enforced by coefficient matching in the monomial basis:
we precompute a fixed linear map $\tilde c = Bc$ such that $p(E)=\sum_{n=0}^{N} \tilde c_n E^n$.
Then coefficient matching yields linear constraints of the form
\begin{equation}\label{eq:sdp_coeff_match_main}
(\tilde c_n - f_n^{(k)}) = \big\langle \Lambda_{n}^{(k)},\, Q^{(k)}\big\rangle,\qquad
k=1,\dots,K,\ \ n=0,\dots,N,
\end{equation}
where $Q^{(k)}$ denotes the appropriate Gram-matrix block
($Q^{(k)}=Q_x^{(k)}\oplus Q_y^{(k)}$ for even $N$, or $Q^{(k)}=Q_z^{(k)}\oplus Q_w^{(k)}$ for odd $N$),
$\langle X,Y\rangle=\mathrm{Tr}[X^\top Y]$, and the fixed symmetric matrices $\Lambda_n^{(k)}$ depend only on the interval endpoints
(and the chosen even/odd certificate). Their explicit construction is given in Appendix~\ref{app:sip_sdp}.

Putting these ingredients together, the SIP \eqref{eq_ub_sip} is equivalent to the following SDP:
\begin{equation}\label{eqn_SDP1}
\begin{aligned}
\underset{c,\{Q^{(k)}\}}{\mathrm{minimize}}\quad & c^\top M\\
\text{subject to}\quad
& (Bc)_n - f_n^{(k)} = \langle \Lambda_{n}^{(k)}, Q^{(k)}\rangle, \quad n=0,\dots,N,\\
& \hspace{2.6em} \forall\, k=1,\dots,K,\\
& Q^{(k)}\succeq 0,\qquad k=1,\dots,K.
\end{aligned}
\end{equation}

The SDP size grows polynomially with $N$ and the number of spectral intervals $K$.
For degree $N$, each interval uses Gram matrices of dimension $\mathcal{O}(N)$, yielding $\mathcal{O}(K N^2)$ scalar decision variables overall.
In practice, moderate degrees ($N\lesssim 20$) and small $K$ are readily handled by standard SDP solvers; see Appendix~\ref{app:sip_sdp} for details.

\subsection{Optimality via duality}

Both the LP (exact-eigenvalue case) and the SDP (interval case) are convex optimization problems.
Convex programs admit associated \emph{dual} problems obtained via Lagrange multipliers; under mild conditions the primal and dual optima coincide (strong duality).
In our setting, the dual variables admit a natural interpretation in terms of \emph{worst-case spectral measures} consistent with the observed moments, yielding both intuition and a direct means to proving optimality.

\paragraph*{Exact-eigenvalue case.}
Consider the dual to \eq{eqn_LP}:
\begin{equation}\label{eqn_LP_dual}
\begin{aligned}
& \underset{\Pi \in \mathbb{R}^{D+1}}{\mathrm{maximize}} & & f^\top \Pi \\
& \text{subject to} & & \mathbf{T}^\top \Pi = M,\qquad \Pi \geq 0.
\end{aligned}
\end{equation}
Here $\Pi$ is a \emph{spectral distribution} over the known eigenvalues, meaning a nonnegative weight vector that defines a discrete measure
\begin{equation}
    d\tilde\mu(E) = \sum_{i=0}^{D} \Pi_i\, \delta(E-E_i)\, dE,
\end{equation}
whose Chebyshev moments match those observed from $\ket{\phi}$:
\begin{equation}
    \sum_{i=0}^{D} \Pi_i\, T_n(E_i) = M_n,\qquad n=0,\dots,N.
\end{equation}
(For $n=0$, this enforces normalization $\sum_i \Pi_i = M_0 = 1$.)
The dual objective $f^\top \Pi$ is the total weight assigned to the indicated eigenvalues.
By strong duality for linear programs, the optimal value of \eq{eqn_LP_dual} equals that of \eq{eqn_LP}, proving that $\mcp_\mcs^+$ is the tightest upper bound that can be inferred from the same finite moment information and spectral constraints (and similarly for the lower bound).

\paragraph*{Interval case.}
The dual program for \eq{eqn_SDP1} can be written as
\begin{equation}\label{eqn_SDP_dual}
\begin{aligned}
    \underset{y}{\mathrm{maximize}}\quad &y^{(0)}_0 \\
    \text{subject to} \quad &\sum_{k=1}^{K} y^{(k)}_n = M_n,\qquad n=0,\dots,N,\\
    &\sum_{n=0}^{N} y^{(k)}_n A^{(k)}_n \preceq 0,\qquad k = 1,\dots,K,
\end{aligned}
\end{equation}
where the matrices $A_n^{(k)}$ are fixed symmetric matrices that encode the linear map from Gram matrices to polynomial coefficients in the Markov--Luk\'acs representation;
their explicit form and the full derivation are given in \Cref{appsec_SIP}.
As in the LP case, the dual constraints enforce (i) moment matching and (ii) physicality of the underlying measure.
Strong duality holds for this SDP~\cite{Lasserre2001}, establishing that the primal bounds produced by \eq{eqn_SDP1} are operationally optimal given the moment information and interval spectral constraints.

\subsection{Obtaining moments in practice}

Computing Hamiltonian moments exactly can incur high costs in both time and memory (see \Cref{appsec_exact_cost}).
For larger systems, one can approximate moments using methods with better system-size scaling, such as matrix product operator/state (MPO/MPS) contractions with controlled bond dimension.
Specifically, approximate moments can be computed via repeated MPO--MPS contractions while truncating the MPS bond dimension by local SVDs during contraction.
We show in \Cref{appsec_mpo_cost} that for a system consisting of $L$ sites with local physical dimension $d$, an MPO bond dimension $D$, and an MPS bond dimension capped at $\chi$, computing the first $N$ moments has leading-order time complexity
\begin{equation}
    C_T \sim \mco(N L d\, D^3 \chi^3),
\end{equation}
and space complexity
\begin{equation}
    C_S \sim \mco(L d\, D^2 + L d^2 \chi^2),
\end{equation}
illustrating practical scalability of the MPO--MPS approach.

When moments are obtained approximately, two issues arise:
(i) the moment sequence may be \emph{non-physical} (no compatible spectral measure), and
(ii) even physical approximate moments may exclude the true measure due to estimation error.
In practice we project approximate moments onto the nearest feasible moment set defined by positive semidefinite moment matrices (Appendix~\ref{app:moment_repair}), and we interpret the resulting behavior of the overlap bounds in the numerical sections below.


\subsection{Toy model: gap dependence and asymmetry of upper and lower bounds}
\label{sec:toy_model}
Before turning to molecular Hamiltonians, we use a controllable toy spectrum to illustrate (i) how certified bounds depend on spectral gaps and (ii) why upper and lower bounds can behave asymmetrically at fixed polynomial degree.
All LP and SDP instances in this section were solved using off-the-shelf convex optimization solvers (primal--dual interior-point methods) with default feasibility and optimality tolerances; the reported bounds are stable at the plotted precision.

\paragraph*{Setup (spectrum, overlaps, and moments).}
Throughout this subsection we work with a normalized spectrum in $[-1,1]$.
Unless stated otherwise, we use $D+1=21$ eigenvalues equally spaced on $[-1,1]$,
\begin{equation}\label{eq:toy_spectrum}
E_i = -1 + 0.1\,i,\qquad i=0,1,\ldots,20,
\end{equation}
so that $E_0=-1$ and $E_D=1$.
We choose a trial-state overlap distribution
\begin{equation}\label{eq:toy_overlaps}
\begin{aligned}
P_0 &= 0.4,\\
P_{i_\pm} &= 0.21 \qquad (E_{i_\pm}=\pm 0.5),\\
P_i &= 0.01 \qquad \text{for the remaining eigenvalues}.
\end{aligned}
\end{equation}
which sums to $1$.
Given $\{E_i,P_i\}$, the moments required by our LP/SDP are computed directly as
\begin{equation}\label{eq:toy_moments}
M_n=\expv{T_n(\hat H_{\rm rs})}_\phi = \sum_{i=0}^{D} P_i\,T_n(E_i),\qquad n=0,\dots,N,
\end{equation}
and the target overlap for $\mcs=\{0\}$ is $P_0$.

\paragraph*{Degree-1 bounds and upper/lower asymmetry.}
We begin with the simplest nontrivial polynomial degree $N=1$.
\Cref{fig:MM} shows the optimal linear majorant/minorant polynomials for $\mcs=\{0\}$.
In this case, the optimal degree-1 upper-bound polynomial is
\bea \label{eqn_linear_ub}
p^{+}_{1}(E) = \frac{E_{D}-E}{E_{D}-E_0}
\quad \Rightarrow \quad
P_0 \le \frac{E_{D}-\expv{\hat H}}{E_{D}-E_0},
\eea
where $\expv{\hat H}=\sum_i P_i E_i$ (here the Hamiltonian is already normalized).
This bound is an ``upper-bound analogue'' of Eckart’s lower bound, and it has the practical advantage that it is always $\le 1$.

\begin{figure}
    \centering
    \includegraphics[width=0.85\linewidth]{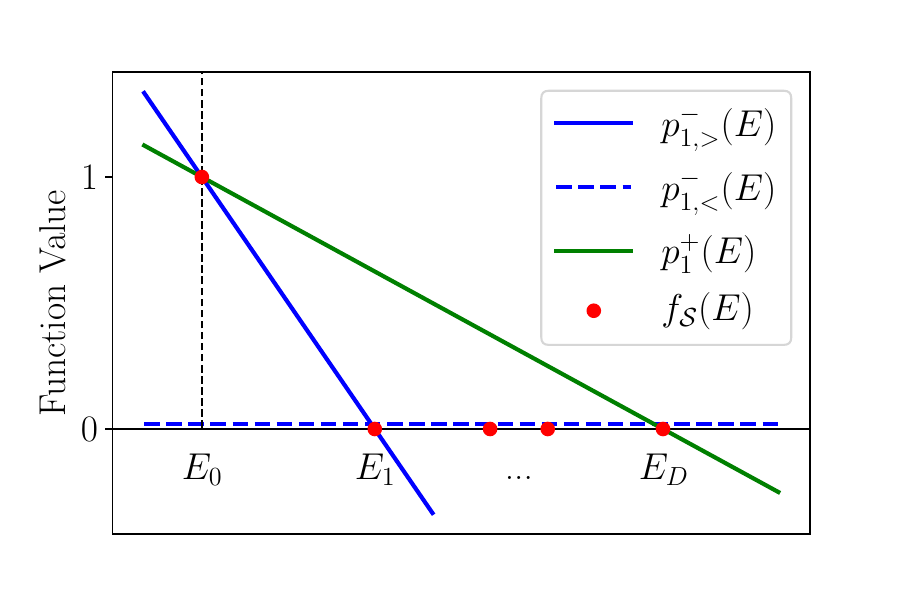}
    \caption{First-order upper (green) and lower (blue) bound polynomials for $\mcs=\{0\}$. For the lower bound, the optimizer selects $p_{1,<}^{-}$ if $E_1<\expv{\hat H}$ (yielding the trivial $0$ lower bound) or $p_{1,>}^{-}$ if $E_1\ge \expv{\hat H}$ (recovering Eckart’s bound).}
    \label{fig:MM}
\end{figure}

To quantify tightness, we use the weighted approximation error
\begin{equation}\label{eqn:error}
    \epsilon_{\mcs}(p_{N}) \coloneqq \sum_{i=0}^{D} P_i \abs{f_{\mcs}(E_i)-p_{N}(E_i)} .
\end{equation}
For the degree-1 upper bound \eqref{eqn_linear_ub}, this error is
\bea\label{eq:UB1_err}
\epsilon_{\mcs}(p^{+}_{1}) = \sum_{i=1}^{D} P_i \frac{E_{D}-E_i}{E_{D}-E_0}.
\eea
It vanishes when all non-target eigenvalues cluster near $E_D$, and it is maximized when they cluster near $E_1$:
\bea\label{eq:UB1w}
\max \epsilon_{\mcs}(p^{+}_{1}) = (1-P_0)\frac{E_D-E_1}{E_{D}-E_0}.
\eea

For the degree-1 lower bound, there are two candidate solutions. 
If $E_1<\expv{\hat H}$, the best certified lower bound is the trivial solution $p^{-}_{1,<}(E)=0$; if $E_1\ge \expv{\hat H}$, the optimizer selects
$p^{-}_{1,>}(E)=(E_1-E)/(E_1-E_0)$, which recovers Eckart’s bound.
Their corresponding errors are
\bea\label{eq:err1}
\epsilon_{\mcs}(p^{-}_{1,<})  = P_0,
\qquad
\epsilon_{\mcs}(p^{-}_{1,>}) = \sum_{i=2}^{D} P_i\frac{E_i-E_1}{E_1-E_0}.
\label{eq:err2}
\eea
Equation~\eqref{eq:err2} shows that nontrivial lower bounds are directly sensitive to the gap $E_1-E_0$ through the denominator.
In particular, the Eckart-type expression can become negative when $E_1<\expv{\hat H}$; our optimization automatically switches to the physically valid $0$ lower bound in that regime.

\paragraph*{Higher degrees: reproducing the optimal polynomials (exact-eigenvalue LP).}
For $N>1$, the optimal upper and lower bounds in the exact-eigenvalue setting are obtained by solving the LP in Sec.~\ref{subsec_exact_EV} (with constraints at the discrete eigenvalues $E_i$).
Concretely, for a chosen degree $N$, one computes moments $M_0,\dots,M_N$ from \eqref{eq:toy_moments}, and then solves:
(i) the upper-bound LP to obtain $p_N^+$ minimizing $c^\top M$ subject to $p_N^+(E_i)\ge f_{\mcs}(E_i)$ for all $i$, and
(ii) the lower-bound LP analog to obtain $p_N^-$ maximizing $c^\top M$ subject to $p_N^-(E_i)\le f_{\mcs}(E_i)$.
\Cref{fig:illustrative} shows the resulting optimal degree-6 polynomials for the toy spectrum \eqref{eq:toy_spectrum} and overlaps \eqref{eq:toy_overlaps}, yielding the certified bound $[0.3692,0.4088]$ for the true value $P_0=0.4$.

\begin{figure}
    \centering
    \includegraphics[width=0.98\linewidth]{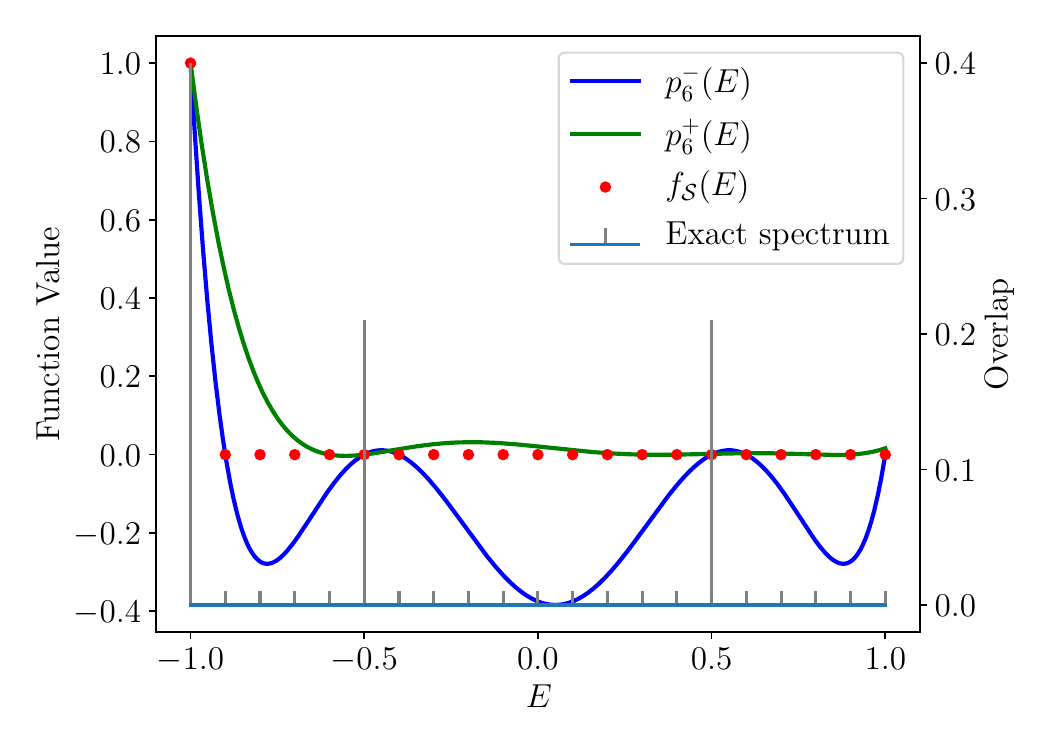}
    \caption{Optimal degree-6 polynomials giving lower/upper bounds on ground-state overlap ($\mcs=\{0\}$) for the toy model \eqref{eq:toy_spectrum}--\eqref{eq:toy_overlaps}, assuming knowledge of exact eigenvalues.}
    \label{fig:illustrative}
\end{figure}

\paragraph*{Gap dependence: reproducing the sweep.}
Equations~\eqref{eq:UB1_err} and \eqref{eq:err2} suggest that upper-bound tightness is governed by the full spectral range $E_D-E_0$, whereas nontrivial lower bounds depend strongly on the gap $E_1-E_0$.
To isolate this effect beyond the linear case, we perform a controlled sweep in which we vary $E_1$ over $(-1,-0.9]$ while keeping all other eigenvalues fixed at \eqref{eq:toy_spectrum} and keeping the overlap weights $\{P_i\}$ fixed.
For each choice of $E_1$ and each polynomial degree $N$, we:
(i) recompute the moments $M_0,\dots,M_N$ via \eqref{eq:toy_moments},
(ii) compute the optimal bounds $p_N^\pm$ by solving the corresponding LPs, and
(iii) evaluate the error $\epsilon_{\mcs}(p_N^\pm)$ via \eqref{eqn:error}.
The results are shown in \Cref{fig:gap_dependence}: the inverse relationship between gap size and lower-bound tightness is clearly observed, and at small gaps the upper bounds typically tighten faster than the lower bounds at the same degree.
This gap sensitivity is an intrinsic limitation of certification from low-order moments and motivates bounding overlap with a \emph{region} (grouping quasi-degenerate states) when appropriate.

\begin{figure}
    \centering
    \includegraphics[width=0.8\linewidth]{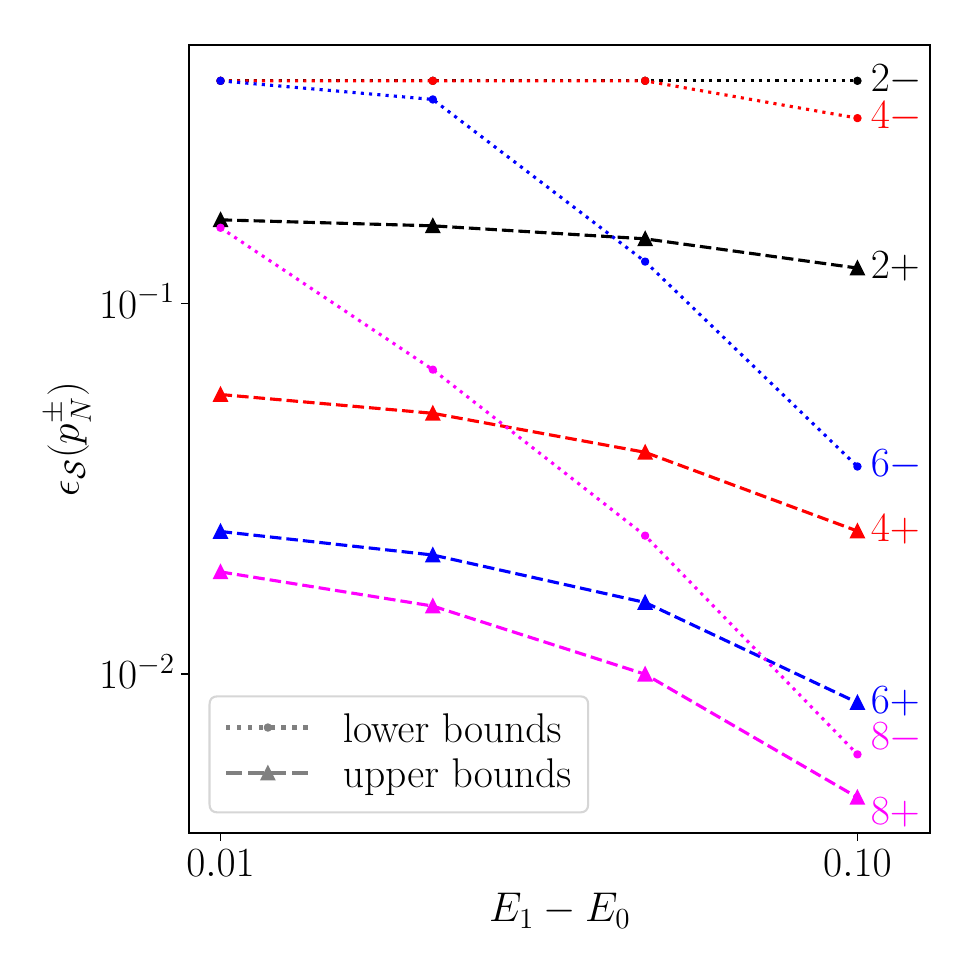}
    \caption{Estimation error for $\mcs=\{0\}$ as a function of the spectral gap $(E_1-E_0)$ in the toy model, for several polynomial degrees (integer before $\pm$). Smaller gaps make certified moment-based bounds intrinsically harder to tighten, especially for lower bounds.}
    \label{fig:gap_dependence}
\end{figure}

\paragraph*{Effect of interval information (interval SDP).}
Finally, we illustrate the additional looseness introduced when only eigenvalue interval enclosures are available.
Using the same toy model as above, we set $\mci=[-1,-0.95)$ and $\ifull=[-1,1]$ and solve the interval SDP \eqref{eqn_SDP1} to obtain the certified bound $[0.3207,0.4426]$, which is looser than the exact-eigenvalue result.
This looseness is expected because the polynomial must satisfy inequality constraints over entire intervals rather than only at discrete eigenvalues: for example, the lower polynomial in \Cref{fig:illustrative} can rise above zero on regions known to be eigenvalue-free, whereas the interval-constrained polynomial in \Cref{fig:illustrative_SIP} must remain below zero over the full complement interval since we do not assume knowledge of where the eigenvalues lie within it.

\begin{figure}
    \centering
    \includegraphics[width=0.98\linewidth]{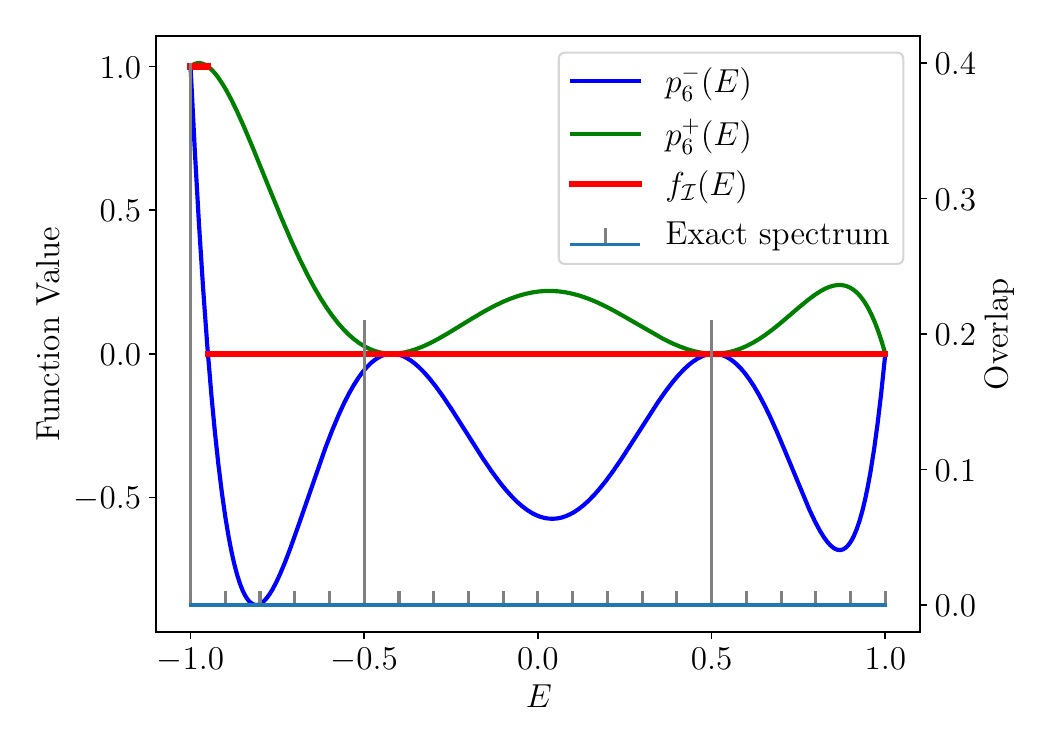}
    \caption{Optimal degree-6 polynomials leading to lower/upper bounds on ground-state overlap for the toy model when only interval enclosures are assumed: $\mci=[-1,-0.95)$ and $\ifull=[-1,1]$.}
    \label{fig:illustrative_SIP}
\end{figure}


\subsection{Strongly correlated molecular Hamiltonians}
\begin{figure*}
    \centering
    \includegraphics[width=0.95\linewidth]{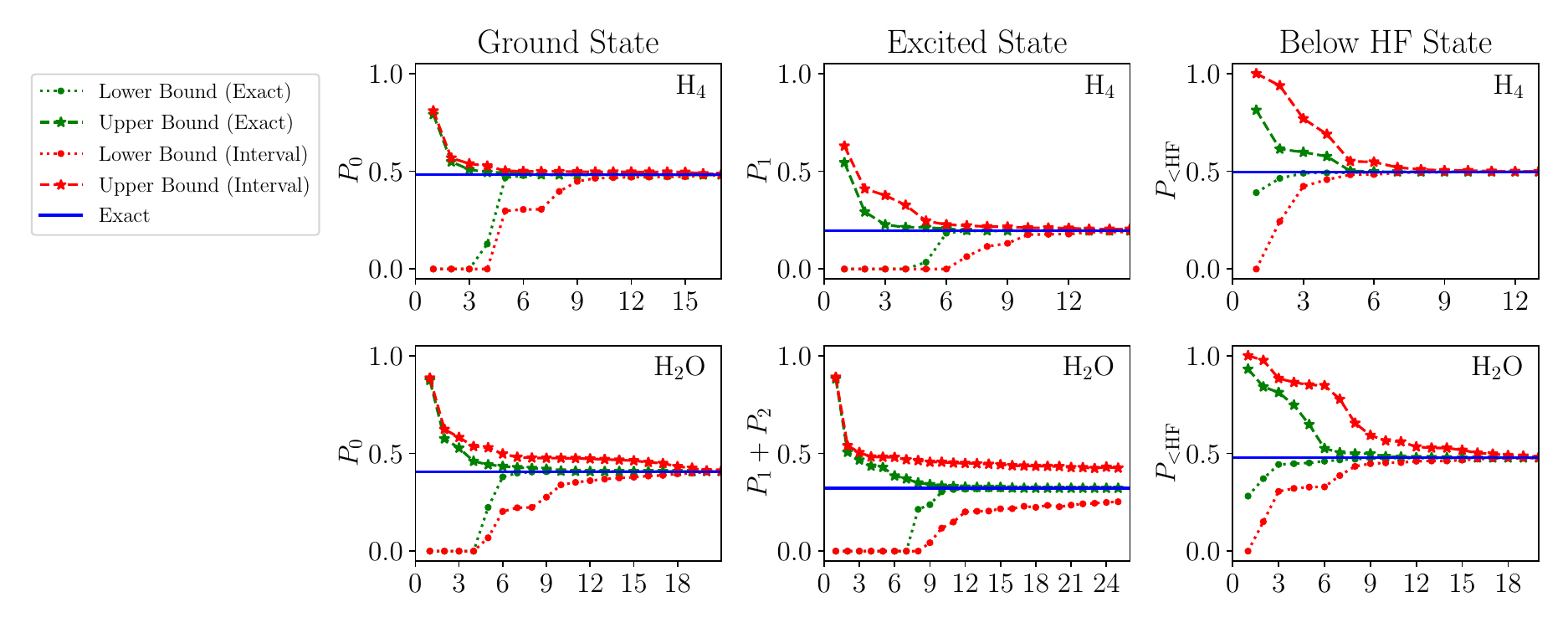}
    \caption{Optimal lower and upper bound values as a function of polynomial degree for two molecular Hamiltonians, computed assuming exact or interval information on the eigenvalue spectrum. Left: overlaps with $\ket{E_0}$. Middle: overlap with $\ket{E_1}$ for H$_4$ or total overlap with $\ket{E_1}$ and $\ket{E_2}$ for H$_2$O. Right: total overlap with all eigenstates below Hartree-Fock energy.
    }
    \label{fig:bounds_vs_polydeg}
\end{figure*}

We assess our algorithms using electronic Hamiltonians for two molecules: H$_4$ and H$_2$O, in the STO-3G basis set.  
Both molecules are arranged at nuclear configurations that correspond to the strongly correlated regime~\cite{Choi2024}.
Full details of Hamiltonian construction and interval formulation are outlined in \Cref{sec_app_mol_details}.
For each system we demonstrate three use cases: when targeting the ground state, selected excited states, or all eigenstates below a particular energy threshold.
The results are shown in \fig{fig:bounds_vs_polydeg}.
The left panel shows the performance of our estimation procedure when targeting $\ket{E_0}$ of each system, with the trial state wavefunction being the Hartree-Fock (HF) state.
It can be observed that the bounds converge to the correct overlap values as the polynomial degree increases.
While the exact eigenvalue bounds always converge faster and set a fundamental upper limit to the performance, the comparable performance of the interval bounds indicates the effectiveness of our method for practical applications.
In particular, the upper-lower bound gaps $\mcp_\mci^+ - \mcp_\mci^-$ are reduced to about 0.2 (H$_4$) and 0.3 (H$_2$O) at polynomial degree 6, and 0.03 (H$_4$) and 0.14 (H$_2$O) at polynomial degree 10 for H$_4$/H$_2$O, demonstrating the effectiveness of the method at low polynomial degrees.
The H$_2$O molecule has a smaller gap for both the ground and excited state scenarios, and therefore requires higher degree polynomials to resolve the overlaps accurately.
This is consistent with the gap dependence highlighted above and illustrates the fundamental constraint of any moment-based method for overlap bounds.
These results highlight several practical takeaways. First, low-degree screening is already informative: already at degree $N=6$--$10$, the interval bounds shrink to gaps of order $0.2$--$0.3$ (degree 6) and $0.03$--$0.14$ (degree 10) in these benchmarks, providing actionable success-probability estimates for pre-QPE screening. Second, interval information is effective: the interval-enclosed bounds track the exact-eigenvalue bounds closely across degrees, indicating that coarse spectral enclosures can be sufficient for practical certification. Finally, spectral gaps control convergence: the slower tightening for H$_2$O compared to H$_4$ is consistent with the gap dependence highlighted by the toy model.

The middle panel shows the performance when targeting selected excited states.
The trial state wavefunction was chosen to be the Slater determinant with the highest overlap with the target eigenstate $\ket{E_1}$. (This determinant was identified using exact diagonalization and is used here only as a controlled benchmark.)
For H$_4$, we computed overlap with the first excited state, and for H$_2$O we computed total overlap with first and second excited state due to their proximity: $E_{2}-E_{1}\approx 1.4\mathrm{mH}$, which is less than chemical accuracy.
For both molecules the bounds converge towards the correct exact overlap values, while the H$_2$O molecule convergence requires higher-degree polynomials due to a smaller energy gap with neighboring states compared to H$_4$.
Finally, the right panel shows bounds for the total overlap $P_{<\mathrm{HF}}$ between HF state and all eigenstates with eigenvalues below the HF energy $E_{\mathrm{HF}}$.
The results show that the obtained bounds readily converge to the exact values in both cases.
Importantly, using only knowledge of $\{\expv{\hat{H}},\expv{\hat{H}^2},\expv{\hat{H}^3}\}$, our algorithm guarantees that there is a significant probability ($\geq 30\%$) for QPE to obtain a lower energy than the classically obtained energy.


\subsection{Approximate moments from MPO--MPS contractions}
Next, to examine the performance of our algorithm where only approximate moments are available, we performed numerical tests using the previously studied active space H$_2$O Hamiltonian, but encoding its $\hat{H}$ as an MPO and using an MPS reference state; computation details are listed in \cref{sec_app_mol_details}.
The resulting MPS state has an overlap of around $0.9924$ with the exact ground state, which is also the only eigenstate with energy lower than the reference energy.
Next, the approximate moments are computed via repeated contraction between the MPO and MPS, while keeping a maximum bond dimension $\chi_{\mathrm{max}}$ of the resulting MPS.
We visualize selected moments in \Cref{fig:inaccuracy_vs_bd} by plotting their relative inaccuracy against the capped bond dimension during contraction.
The relative inaccuracy is defined by 
\begin{equation}
    \epsilon_{\mathrm{MPS}} = \frac{\abs{M_n^{\mathrm{MPS}}-M_n^{\mathrm{exact}}}}{\abs{M_n^{\mathrm{exact}}}},
\end{equation}
where the exact moments are calculated without truncation using a converted matrix-vector format.

\begin{figure}
    \centering
    \includegraphics[width=0.98\linewidth]{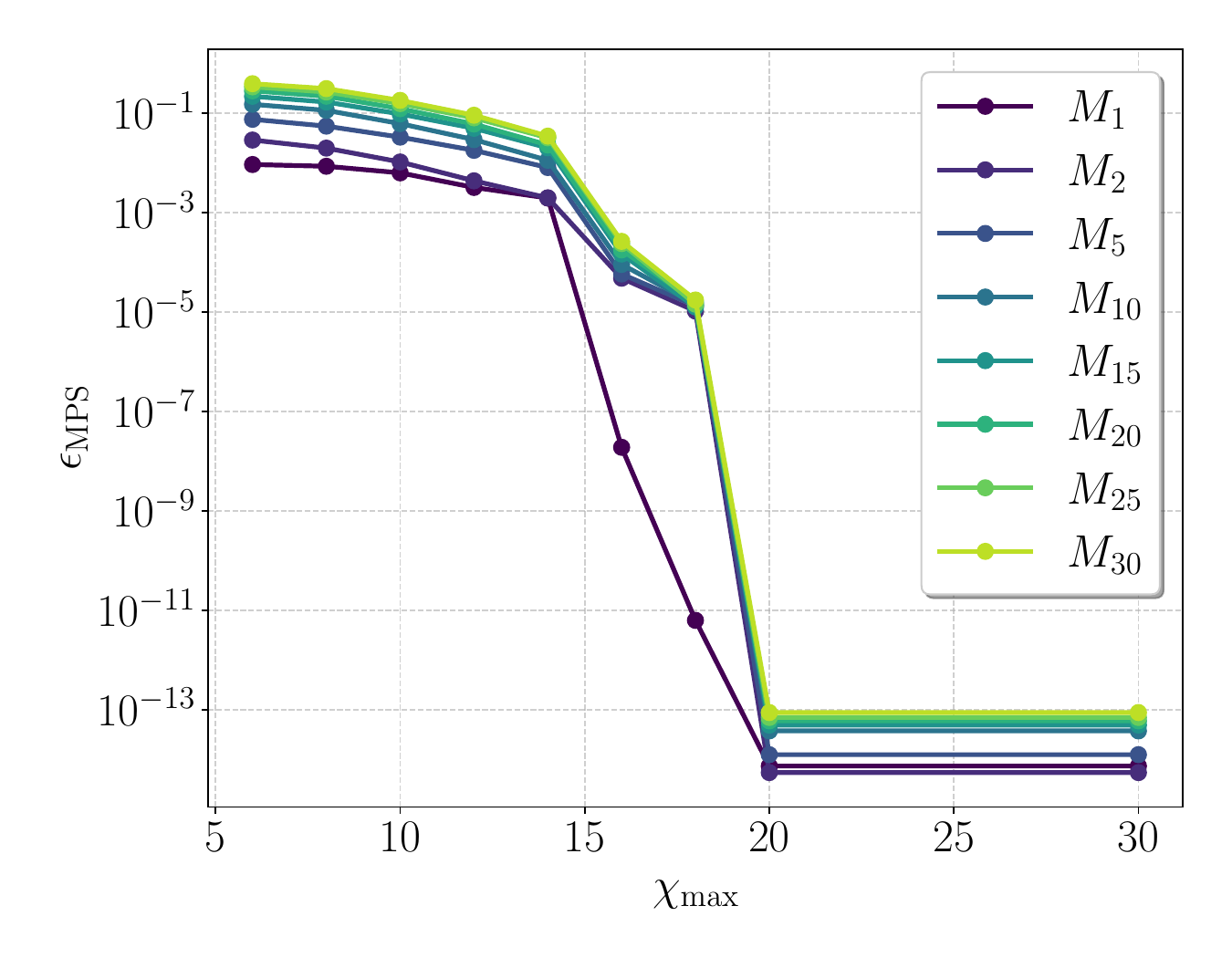}
    \caption{Relative inaccuracy of selected moment terms computed using MPO--MPS contraction with bond dimensions capped at $\chi_{\mathrm{max}}$, for the active-space H$_2$O Hamiltonian used in this section.}
    \label{fig:inaccuracy_vs_bd}
\end{figure}

\fig{fig:inaccuracy_vs_bd} shows an empirically close-to-exponential improvement in accuracy with increasing bond dimension cap for all the moments, and for this instance there is a cutoff bond dimension around 20, beyond which all the moments become exact.
We also observe the trend that higher moments have lower accuracy as expected, yet the degree of inaccuracy remains bounded as the moment order increases and tends to saturate.
These trends suggest that for larger systems, accurate moments could be computed with a practically feasible $\chi_{\mathrm{max}}$.

We then simulate the effects of truncated bond dimensions in practice using the previously computed moments for the H$_2$O molecule, with results shown in \fig{fig_multi_bd_comp}. 
For each capped bond dimension, \eq{eqn_optimized_moments} with $\lambda=10^6$ and $\epsilon$ starting at $10^{-18}$ was used to obtain the physically meaningful $M_{\mathrm{op}}$ vectors closest to the raw moments. 
Then, lower/upper bounds for overlap with all states below MPS reference energy are computed using $M_{\mathrm{op}}$'s.
For moments computed with low bond dimensions (particularly 6 and 8), they deviate significantly from the exact moments, such that all compatible probability measures have very low overlap with the true ground state.
As a result, the bounds converged to incorrectly low overlap values.
Meanwhile, the moments computed with higher bond dimensions are closer to the correct values, which is reflected in the correctly converged bounds at high polynomial degrees that approach the exact overlap value.

\begin{figure}
    \centering
    \includegraphics[width=0.98\linewidth]{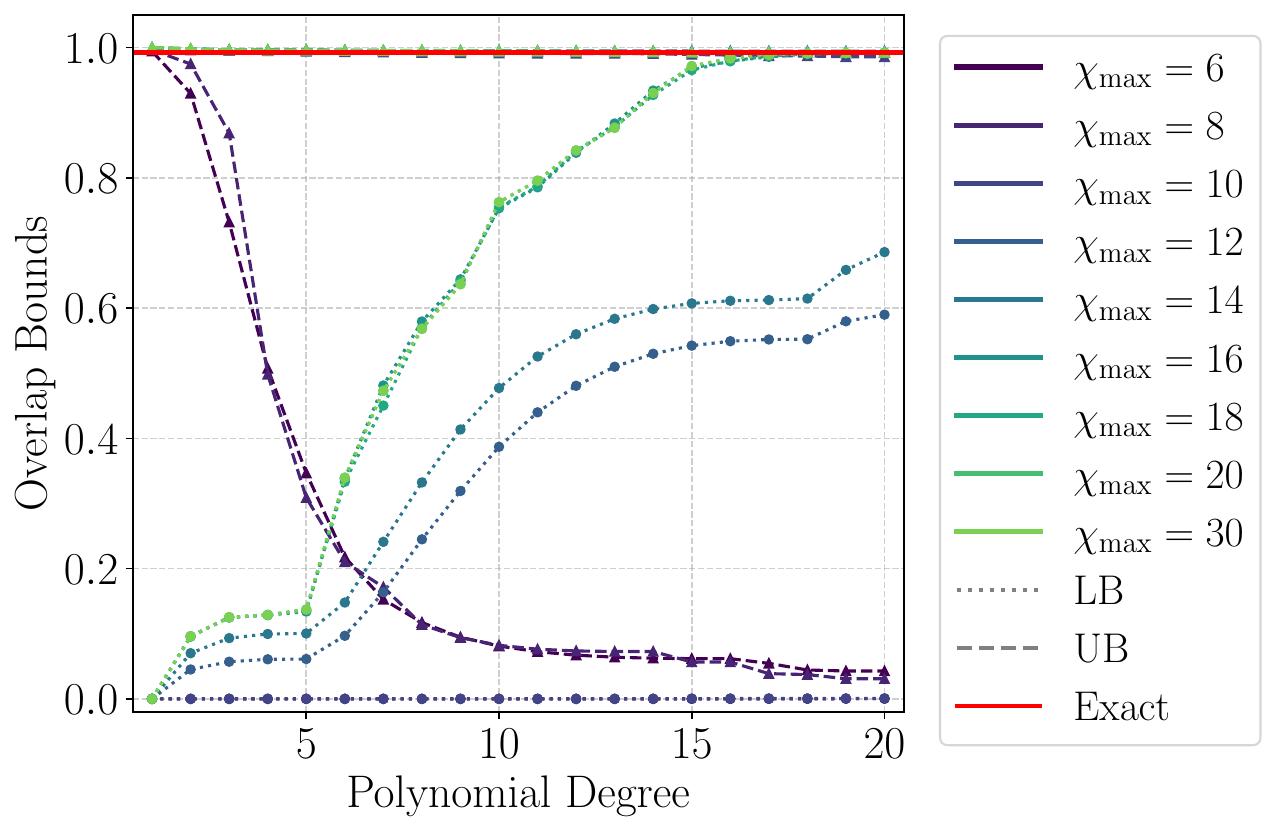}
    \caption{Optimal lower/upper bounds on overlap between MPS state and all eigenstates below its energy, assuming different $\chi_{\mathrm{max}}$  used in MPO-MPS contraction during approximate moment calculation. 
    }
    \label{fig_multi_bd_comp}
\end{figure}

\section{Conclusions}
 
We have presented a classical algorithm that computes systematically improvable lower and upper bounds on the overlap between an input state and arbitrary Hamiltonian eigenstates, or all eigenstates within a specified spectral region. 
Our method requires only Hamiltonian-moment expectation values and coarse spectral information (exact eigenvalues or interval enclosures), and it returns systematically improvable upper and lower bounds on target overlaps.
We established optimality of the computed bounds using linear/semidefinite programming theories, showing that they are indeed the optimal values among all worst-case spectral measures compatible with the moments.
Compared to previous approaches, our framework provides certified bounds that are systematically improvable and (in our benchmarks) tighten rapidly with polynomial degree, while remaining computationally feasible for a range of strongly correlated molecular systems.
Similar to the Eckart bounds, our results show that the tightness of moment-based bounds deteriorates as the relevant spectral gap decreases, reflecting a fundamental limitation shared by overlap estimates based solely on low-order moments.
We also studied the scenario where only approximate moments can be obtained due to limited computational resources, and discussed the limitations and remedies under such cases. A natural next step is to incorporate explicit moment-uncertainty models (e.g., confidence intervals for $M_n$) and compute robust worst-case bounds over those uncertainty sets, thereby preserving certification guarantees under noisy moment estimates.
Additionally, we note that expectation values of Chebyshev polynomials of the Hamiltonian can be efficiently obtained on a quantum computer using qubitization and block-encoding techniques~\cite{Low2019}. 
This synergy offers a natural pathway for hybrid classical-quantum strategies, where classical computations serve as a screening tool while quantum computers provide the final, high-accuracy results.

A key limitation of our approach arises when the energy gap between neighboring eigenstates is small, reducing the effectiveness of Hamiltonian power-based bounds. 
Incorporating Hamiltonian symmetries by preparing symmetry-adapted initial states can mitigate this issue. 
However, in cases of quasi-degenerate states within the same symmetry sector, any method relying on Hamiltonian expectation values faces inherent challenges. 
Our solution is to estimate the total overlap with the quasi-degenerate eigen-subspace rather than individual eigenstates, as done in the H$_2$O excited state molecular example.

In large systems, however, this approach may still face difficulties. Even if multiple approximate states significantly overlap with the quasi-degenerate subspace, they may fail to resolve all individual eigenstates. For instance, if two approximate states $\ket{\phi_0}$ and $\ket{\phi_1}$ collectively have 10\% overlap with a quasi-degenerate subspace spanned by $\ket{\psi_0}$ and $\ket{\psi_1}$, they may still have negligible overlap with
$\ket{\psi_0}$ or $\ket{\psi_1}$ individually, making individual eigenstates inaccessible. Orthogonality among approximate states can be maintained by contributions from other eigenstates outside the quasi-degenerate manifold.

An additional key future direction lies in developing more robust protocols to estimate Hamiltonian moments.
For MPO-based methods, a solid theoretical framework on how errors scale with the bond dimension would be very useful.
Another possible direction is to obtain increasingly more accurate classical energy estimates and/or initial states alongside the moment calculations, without incurring additional cost.
As an alternative to the primitive method in \eq{eqn_optimized_moments}, better-motivated moment optimization that accounts for errors in different moment orders could also enhance the interpretability of computed bounds.

\begin{acknowledgments}
The authors thank Thomas Ayral, Seonghoon Choi, Joshua Cantin, Ignacio Loaiza, and Robert A. Lang for helpful discussions. This work was funded under the DARPA IMPAQT program, under grant No. HR0011-23-3-0028. This research was partly enabled by Compute Ontario (computeontario.ca) and the Digital Research Alliance of Canada (alliancecan.ca) support. Part of the computations were performed on the Niagara supercomputer at the SciNet HPC Consortium. SciNet is funded by Innovation, Science, and Economic Development Canada, the Digital Research Alliance of Canada, the Ontario Research Fund: Research Excellence, and the University of Toronto.

\end{acknowledgments}

\appendix
\section{Solving semi-infinite program via Markov-Lukács certificates}
\label{app:sip_sdp}
\label{appsec_SIP}
Here we describe how a semi-infinite program (SIP) involving polynomial constraints can be solved by applying the method of Lukács certificate of nonnegativity. 
These results allow univariate SIPs to be reformulated as a semidefinite program (SDP) and can be solved using standard SDP solvers.
Recall that the SIP for obtaining the overlap upper bound takes the form
\begin{equation}\label{eqn_app_sip}
\begin{aligned}
    \underset{c \in \mathbb{R}^{N+1}}{\mathrm{minimize}} \quad &c^\top M  \\ \text{subject to } \quad &p(x) - f_{\mci}(x) \ge 0, \;\; \forall x \in \ifull.
\end{aligned}
\end{equation}
Here we only demonstrate with the above upper bound problem in \eqref{eqn_app_sip} and the lower bound case generalizes trivially.
To turn \eq{eqn_app_sip} into a more easily solvable form, we use algebraic certificates of nonnegativity.
For the univariate case, the Lukács theorem provides an exact certificate of nonnegativity on an interval. 
Specifically, depending on the even/odd parity of $N$, a real polynomial $q_{\mathrm{nn}}(t)$ of degree $N$ is non-negative on the interval $[-1,1]$ if and only if it admits a representation~\cite{Szeg1939}
\begin{equation}
    q_{\mathrm{nn}}(t) = \begin{cases}
        \; a(t)^2 + (1-t^2)\, b(t)^2, \quad N=2m \\
        \; (t+1)\, c(t)^2 \;+\; (1-t)\, d(t)^2, \quad N=2m+1
    \end{cases}
\end{equation}
where $m \in \mathbb{N}^+$ and $(a,b,c,d)$ are polynomials with degrees:
\begin{equation*}
    \deg a \le m,\ \deg b \le m-1,\ \deg c \le m,\ \deg d \le m.
\end{equation*}
Note the distinction between even and odd degrees: in the even case, the factor $1-t^2$ vanishes at both endpoints, while in the odd case the endpoint factors $1\pm t$ appear separately.

By affine rescaling, we can extend the above result for any finite interval $[l,r]$, using the affine map
\begin{equation}
    t(s) = \frac{2x-(l+r)}{r-l} \Longleftrightarrow
x(t) = \frac{(r-l)t + l+r}{2}.
\end{equation}
which bijects $[l,r]$ with $[-1,1]$.
Applying this to the original problem, we get for even degrees:
\begin{equation}
    q_{\mathrm{nn}}(s) = [a(t(s))]^2 + (s-l)(r-s)[\frac{2b(t(s))}{r-l}]^2
\end{equation}
and for odd degrees:
\begin{equation}
    q_{\mathrm{nn}}(s) = \frac{2(x-l)}{r-l}[c(t(s))]^2 + \frac{2(r-x)}{r-l}[d(t(s))]^2.
\end{equation}
In both cases, the positive scalar factors $\tfrac{4}{(r-l)^2}$ or $\tfrac{2}{r-l}$ can be absorbed into the squares. 
We thus obtain the following generalized version of Lukács theorem for a variable $s\in [l,r]$: a real polynomial $q_{\mathrm{nn}}(s)$ of degree $N$ is nonnegative on the interval $[l,r]$ if and only if it admits a representation
\begin{equation}\label{eqn_app_qnn}
    q_{\mathrm{nn}}(s) = \begin{cases}
        \; a(s)^2 + (s-l)(r-s)\, b(s)^2, \quad N=2m \\
        \; (x-l)\, c(s)^2 + (r-x)\, d(s)^2, \quad N=2m+1
    \end{cases}
\end{equation}
where $m \in \mathbb{N}^+$ and $(a,b,c,d)$ are polynomials with degrees:
\begin{equation*}
    \deg a \le m, \deg b \le m-1,\deg c \le m, \deg d \le m.
\end{equation*}

\cref{eqn_app_qnn} thus gives the exact interval certificates on $[l,r]$.
We will use these to enforce positivity constraints.
Specifically, to ensure $p_c(s) - f(s) \ge 0$ for all $x \in [l,r]$, we enforce that it has the form of \cref{eqn_app_qnn}.
To linearly parametrize the squared polynomials, we use another fact that a polynomial $a(s)$ is a perfect square if and only if there exists a positive semidefinite (PSD) matrix $Q_a$ such that
\begin{equation}\label{eqn_app_asquare}
    a(s)^2 = v(s)^\top\, Q_a \, v(s),
\end{equation}
where $v(s)$ is the standard monomial basis $v(s)=[1,s,s^2,\dots]^\top$ and $Q_a$ is the Gram matrix of polynomial $a$.
The other squared polynomials have exactly the same parametrization.
This allows us to linearly parametrize $q_{\mathrm{nn}}(t)$ by $Q_{a}\, \oplus \, Q_{b}$ for even $N$, or $Q_{c}\, \oplus \, Q_{d}$ for odd $N$.

Next, since nonnegativity must hold on each of the $K$ distinct intervals of $\ifull$ on which the piecewise $f_{\mci}(E)$ is defined, we need distinct certificates over each interval $k$, parametrized by $Q_{a}^{(k)} \dots Q_{d}^{(k)}$.
This leads to the following SDP whose solution equals that of the desired SIP:
\begin{equation}\label{eqn_app_SDP1}
\begin{aligned}
    \underset{c \in \mathbb{R}^{N+1}}{\mathrm{minimize}} \quad &c^\top M  \\ \text{subject to} \quad &p(E) - f_{\mci}^{(k)}(x) = q_{\mathrm{nn}}^{(k)}(x)\\
    &Q^{(k)} \succeq 0,\  k = 1\dots K,
\end{aligned}
\end{equation}
where we used $Q^{(k)}$ to denote $Q_{a}^{(k)}\oplus Q_{b}^{(k)}$ for even $N$, or $Q_{c}^{(k)} \oplus  Q_{d}^{(k)}$ for odd $N$.

To reformulate the above SDP into the standard form, recall the definitions
\begin{equation}
    p_{c}(x)=\sum_{n} c_n x^n,\ f^{(k)}(x)=\sum_{n} f_n^{(k)} x^n.
\end{equation}
By matching the coefficients of each $x$ degree, the equality constraint in \eq{eqn_app_SDP1} can be re-formulated as:
\begin{equation}\label{eqn_app_coeff_match}
\begin{aligned}
    c_n - f_n^{(k)} &= \langle A_n^{(k)}, Q_{a}^{(k)} \rangle + \langle B_n^{(k)}, Q_{b}^{(k)} \rangle \\
    &= \langle \Lambda^{(k)}_{n}, Q^{(k)} \rangle \quad \forall n=0\ldots N
\end{aligned}
\end{equation}
with the standard definition of matrix norm, $\langle X,Y \rangle=\Tr[X^\top Y]$, and defining $\Lambda^{(k)}$ as $A_{n}^{(k)}\oplus B_{n}^{(k)}$ for even $N$, or $C_{n}^{(k)} \oplus  D_{n}^{(k)}$ for odd $N$.

The matrices $A_n^{(k)} \dots D_n^{(k)}$ are fixed symmetric matrices converting the linear map from Gram matrices to polynomial coefficients in the Markov–Lukács representation.
The explicit forms of such matrices are left to the end of this section.
We have thus converted \Cref{eqn_app_SDP1} into the following standard SDP form:
\begin{equation}\label{eqn_app_SDP2}
\begin{aligned}
    \underset{c \in \mathbb{R}^{N+1}}{\mathrm{minimize}} \quad &c^\top M \\ \text{subject to} \quad &c_n - f_n^{(k)} = \langle \Lambda^{(k)}_{n}, X^{(k)} \rangle\\
    &X^{(k)} \succeq 0 \\ 
    & \forall k = 1\dots K,\ n=0 \ldots N.
\end{aligned}
\end{equation}

Next we compute the dual form of \Cref{eqn_app_SDP2}.
The variables $X^{(k)}$ are constrained to be in a PSD cone.
Thus, we apply the Karush–Kuhn–Tucker (KKT) theory~\cite{Yang2017} which generalizes the standard Lagrangian theory.
Define the dual variables $y^{(k)}_n$ for each power $n$ and constraint region $k$.
Using them to construct the Lagrangian:
\begin{equation}
    \mathcal{L}=c^\top M + \sum_{n,k} y^{(k)}_n (f^{(k)}_n - c_n + \langle \Lambda^{(k)}_{n}, X^{(k)} \rangle).
\end{equation}
For the free variables $c_n$, the stationary conditions are
\begin{equation}
    \frac{\partial \mathcal{L}}{\partial c_n} = M_n - \sum_{k} y^{(k)}_n =0 \Rightarrow \sum_{k} y^{(k)}_n = M_n.
\end{equation}
For the constrained variables $X$, the stationary conditions are
\begin{equation}
    \frac{\partial \mathcal{L}}{\partial X^{(k)}} \preceq 0 \Rightarrow \sum_{n=0}^{N} y^{(k)}_n A^{(k)}_n \preceq 0.
\end{equation}
Physically the above condition implies that no PSD perturbation of $X^{(k)}$ can improve the Lagrangian.
Thus the dual program for computing the upper bound is given by
\begin{equation}\label{eqn_app_SDP_dual}
\begin{aligned}
    \underset{y}{\mathrm{maximize}} \quad &\sum_{k,n} y^{(k)}_n f^{(k)}_n \\ \text{subject to} \quad &\sum_{k=1}^{K} y^{(k)}_n = M_n,\ n=0 \ldots N\\
    &\sum_{n=0}^{N} y^{(k)}_n A^{(k)}_n \preceq 0,\ k = 1\dots K
\end{aligned}
\end{equation}
And because $f_\mci$ is either $1$ or $0$ for our purpose, the only non-zero term left in the maximization objective is $y_0^{(0)}$, simplifying the expression to
\begin{equation}
\begin{aligned}
    \underset{}{\mathrm{maximize}}\quad &y^{(0)}_0 \\ \text{subject to} \quad &\sum_{k=1}^{K} y^{(k)}_n = M_n,\ n=0 \ldots N\\
    &\sum_{n=0}^{N} y^{(k)}_n A^{(k)}_n \preceq 0,\ k = 1\dots K
\end{aligned}
\end{equation}
and similarly for the lower bound.

We now present the explicit form of the symmetric conversion matrices. They are obtained via collecting the polynomial terms with the same powers and writing the collected coefficients in matrix form.
For example, expanding each term in \eq{eqn_app_asquare}:
\begin{equation}
    v(s)^\top\, Q^{(k)}_a \, v(s) = \sum_{i,j=0}^{m} (Q^{(k)}_a)_{ij} x^{i+j}
\end{equation}
shows that the coefficient of $x^n$ is $\sum_{i+j=n} (Q^{(k)}_a)_{ij}$.
This can be written in matrix inner product form as
\begin{equation}
    \sum_{i+j=n} (Q^{(k)}_a)_{ij} = \langle A^{(k)}_n, Q^{(k)}_a \rangle
\end{equation}
with
\begin{equation}
    (A^{(k)}_n)_{ij} = \begin{cases}
        1, &i+j=n \\
        0, &\mathrm{otherwise.}
    \end{cases}
\end{equation}
The symmetric conversion matrices for the Gram matrices $Q^{(k)}_{b}$, $Q^{(k)}_{c}$, $Q^{(k)}_{d}$ can all be constructed via the same argument.
The only difference is that instead of ``raw'' polynomials in the $Q^{(k)}_{a}$ case, the Gram-polynomials are multiplied by an additional polynomial in the final form of $q_{\mathrm{nn}}(s)$, which modifies the resulting coefficients for each $x$ power and thus the form of the conversion matrices.
We include them below for completeness:
\begin{equation}
    (B^{(k)}_n)_{ij} = \begin{cases}
        -l^{(k)}r^{(k)}, &i+j=n \\
        r^{(k)}+l^{(k)}, &i+j=n-1 \\
        -1, &i+j=n-2\\
        0, &\mathrm{otherwise.}
    \end{cases}
\end{equation}
\begin{equation}
    (C^{(k)}_n)_{ij} = \begin{cases}
        -l^{(k)}, &i+j=n \\
        1, &i+j=n-1 \\
        0, &\mathrm{otherwise.}
    \end{cases}
\end{equation}
\begin{equation}
    (D^{(k)}_n)_{ij} = \begin{cases}
        r^{(k)}, &i+j=n \\
        -1, &i+j=n-1 \\
        0, &\mathrm{otherwise.}
    \end{cases}
\end{equation}
In general, when multiplied by a polynomial, the skew-diagonals of the resulting matrix are modified according to the multiplied polynomial.

\section{Complexity of Hamiltonian Moment Evaluation Using Truncated MPO–MPS Propagation}
\label{appsec_mpo_cost}
We analyze the computational complexity of evaluating the first $N$ Hamiltonian moments,
\begin{equation}
    \bra{\phi} \hat{H}^n \ket{\phi},~n=1\dots N,
\end{equation}
using an iterative MPO–MPS propagation scheme with fixed maximum bond dimension truncation at every step.

The Hamiltonian $\hat{H}$ is represented as a matrix product operator (MPO) with maximum bond dimension $\chi$, and the wavefunction is represented as a matrix product state (MPS) with maximum bond dimension $B$. 
The system consists of $L$ sites, each with local physical dimension $d$.
Here $L$ would be equal to either the number of spatial or spin orbitals, and $d$ would either be $4$ or $2$ correspondingly.
At each iteration, the algorithm applies the MPO to the current MPS and subsequently truncates the resulting state by performing a local singular value decomposition (SVD) at every bond to enforce a maximum bond dimension 
$B$.
The cost for the two main operations are: 
\begin{enumerate}
    \item Cost of a single MPO-MPS application.
    After contracting the MPO and MPS tensors at a given site, the intermediate bond dimension grows from $B$ to approximately $B\chi$. The local tensor contraction cost scales as $\mco(d^2 B^2 \chi^2)$. 
    Summed over all sites, the total contraction cost is
    \begin{equation}
        \mco(Ld^2 B^2 \chi^2).
    \end{equation}
    \item Cost of truncation.
    Following contraction, truncation is performed locally at each bond via an SVD of an effective matrix of dimension $(B\chi d) \times (B\chi)$. The cost of a single SVD scales as $\mco((B\chi d)(B\chi)^2)=\mco(dB^3 \chi^3)$. 
    Since truncation is performed at every site, the total truncation cost is
    \begin{equation}
        \mco(LdB^3 \chi^3).
    \end{equation}
    Since the physical or typical DMRG regimes with moderately large $B$, the truncation cost dominates the contraction cost.
\end{enumerate}

Each application of $\hat{H}$ followed by truncation therefore has a leading order cost of $\mco(LdB^3 \chi^3)$.
Since the algorithm constructs $N$ successive states $\ket{\phi_n} \approx \hat{H}^n \ket{\phi}$, the total time complexity for computing the first $N$ moments scales as
\begin{equation}
    \mco(NLdB^3 \chi^3),
\end{equation}
which is the leading order cost since computing the final inner product $\braket{\phi}{\phi_n}$ scales as $\mco(L d B^2)$ and is subleading.

To estimate the total space complexity, note that at any time, the algorithm stores a constant number of MPSs and a single MPO. 
The memory cost therefore scales as
\begin{equation}
    \mco(L d B^2 + L d^2 \chi^2),
\end{equation}
which is independent of the total moment $N$.

\subsection{Moment physicality repair for approximate moments}
\label{app:moment_repair}
When moments are computed approximately (e.g., via MPO--MPS contractions with truncation), the resulting sequence may violate basic positivity conditions required of a valid spectral measure. 
To obtain a physically consistent moment sequence, we project the raw moments onto the nearest feasible set defined by positive semidefinite (PSD) moment matrices.

Given raw moments $M_0,\ldots,M_N$, define the Hankel moment matrix $G(M)$ and the shifted (Stieltjes-type) matrix $G_S(M)$ by
\begin{equation}
  [G(M)]_{ij}=M_{i+j}, \qquad [G_S(M)]_{ij}=M_{i+j}-M_{i+j+2},
\end{equation}
with indices $i,j$ chosen so that the required moments exist (here we take the largest square matrices compatible with order $N$).
We then compute an optimized sequence $M_{\mathrm{op}}$ by solving
\begin{equation}\label{eqn_optimized_moments}
\begin{aligned}
 \underset{M_{\mathrm{op}}\in\mathbb{R}^{N+1},\, s\ge 0}{\mathrm{minimize}}\quad &\sum_{n=1}^{N} \left(M_{\mathrm{op},n}-M_n\right)^2 + \lambda s \\
 \text{subject to}\quad & G(M_{\mathrm{op}})+ s I \succeq \epsilon I,\\
 & G_S(M_{\mathrm{op}})+ s I \succeq \epsilon I.
\end{aligned}
\end{equation}
Here $s$ is a slack variable that preserves feasibility when the raw moments are substantially nonphysical, $\lambda$ is a large penalty parameter prioritizing feasibility, and $\epsilon$ is a small stability shift. 
In Sec.~III we use $M_{\mathrm{op}}$ to compute overlap bounds and illustrate how insufficient bond dimension can still lead to physically consistent but biased moments, yielding incorrect bounds at high polynomial degree.

\section{Complexity of Hamiltonian Moment Evaluation in the Determinant Basis}
\label{appsec_exact_cost}
We analyze the computational complexity of evaluating Hamiltonian moments
\begin{equation}
    \bra{\phi} \hat{H}^n \ket{\phi},~n=1\dots N,
\end{equation}
when the Hamiltonian $H$ is represented in operator form and the reference state $\ket{\phi}$ is stored explicitly as a sparse vector in the many-body determinant basis.
The system consists of $N_o$ spin orbitals and $N_e$ electrons, with Hilbert space dimension
\begin{equation}
    D_{\mathrm{hs}} = \binom{N_o}{N_e}.
\end{equation}
The Hamiltonian includes one- and two-body terms and is applied using Slater–Condon rules without forming a dense matrix.

Applying the Hamiltonian to a single determinant produces $\mco(N_o^4)$ connected determinants due to the two-body interaction. 
If a state contains $S$ nonzero determinants, applying $\hat{H}$ a new state with at most $\mco(S N_o^4)$ nonzero components. 
Iterating this process,
\begin{equation}
    S_n \sim \mco(S_0 N_o^{4n})
\end{equation}
until saturation at the full Hilbert space dimension $D_{\mathrm{hs}}$.
Note that if the initial state contains additional symmetry besides electron number, the final size could be smaller than $D_{\mathrm{hs}}$.
\begin{enumerate}
    \item Time complexity: applying the Hamiltonian to a state with $S$ nonzero determinants costs
    \begin{equation}
        \mco(S N_o^4).
    \end{equation}
    The cost of constructing the $n$-th moment state $\ket{\phi_n} = \hat{H}^n \ket{\phi}$ therefore scales as
    \begin{equation}
        \mco(S_0 N_o^{4n}).
    \end{equation}
    Summing over the first $N$ moments, the total time complexity is
    \begin{equation}
        \mco(S_0 N_o^{4N})
    \end{equation}
    until saturation, after which it becomes $\mco{ND_{\mathrm{hs}}N_o^4}$
    \item Space complexity: The dominant memory requirement is the storage of the propagated state $\ket{\phi_n}$, which contains $S_n$ nonzero determinants. 
    Thus, the space complexity scales as
    \begin{equation}
        \mco(\min(S_0 N_o^{4N},D_{\mathrm{hs}})).
    \end{equation}
\end{enumerate}

\section{Details of Molecular Simulations}
\label{sec_app_mol_details}
Here we provide the technical details on the settings used in the molecular simulation examples.
For the molecular geometries, H$_4$ is taken in the linear configuration with bond length $2.0$ \AA, and 
H$_2$O is at $R({\rm O-H}) = 2.1$ \AA\ and H-O-H angle $107.6 \degree$. 
For the H$_2$O molecule, an active space model is used where the 3 lowest occupied Hartree-Fock spatial orbital is frozen; and for the H$_4$ molecule, the full Hilbert space is used.

Next we describe the exact and interval indicator functions used in both H$_4$ and H$_2$O molecular simulations.
For both the ground-state and excited-state estimation tasks, for $f_\mcs(E)$ we set $\sfull$ to include all eigenvalues with corresponding overlaps above $10^{-20}$.
We used $\mcs = \{0\}$ for the ground-state tasks, and $\mcs = \{1\}$ for the H$_4$ excited-state task and $\mcs = \{1,2\}$ for the H$_2$O excited-state task.
For $f_\mci(E)$, we set $\ifull=[E_L,E_U]$ and 
\begin{equation}\label{eqn_simulation_interval}
\begin{aligned}
\mci=[E_i-\gamma^- g_{i-1}, E_i+\gamma^+ g_{i}]
\end{aligned}
\end{equation}
where we used $\gamma^{\pm}=0.3$ as the lower (upper) scale factor and $g_i=E_{i+1}-E_i$ is the energy gap.
Both the Hamiltonian and the energy intervals are then rescaled with $E_L$/$E_U$ set to $E_0$ and $E_D$ rounded down/up to the closest decimal (see \eq{eqn:rescaled_H}) to emulate approximate spectral lower/upper bounds.
Finally, for the ``below HF'' experiment we set $\mci=[-1,E_{\mathrm{HF}})$ and $\ifull=[-1, 1]$.

For the MPO-MPS experiment with H$_2$O molecule, using the Pyblock3~\cite{Zhai2021} package, we first encoded the Hamiltonian operator into an MPO as a pyblock3.hamiltonian.Hamiltonian object.
Then, we initialized an MPS state using pyblock3.hamiltonian.Hamiltonian.build\_mps with bond\_dim=12.
The size of the ansatz MPS was kept small to avoid obtaining the exact ground state.
To produce the DMRG state we used the pyblock3.algebra.mpe.MPE.dmrg function with 2 DMRG sweeps, a noise scheme starting as $10^{-6}$ and goes to $0$, and a Davidson threshold of $10^{-10}$ (n\_sweeps=2, noises=[1e-6,0], tol=1e-10).

\bibliographystyle{apsrev4-2}
\bibliography{main}
\end{document}